\documentstyle[11pt,aas2pp4,tighten,flushrt,epsf]{article}

\begin{document}

\title{Counter-Rotating Accretion Disks} 
\author{R.V.E. Lovelace}
\affil{Department of Astronomy, Cornell University, Ithaca, NY 14853}
\author{Tom Chou}
\affil{Laboratory of Atomic and Solid State Physics, Cornell University,
Ithaca, NY 14853}
\date{\today}

\begin{abstract}
We consider accretion disks consisting
of counter-rotating gaseous components
with an intervening shear layer.
Configurations of this type may arise
from the accretion of newly supplied
counter-rotating gas onto an existing
co-rotating gas disk. For simplicity we consider 
the case where the gas well above the disk
midplane is rotating with angular rate
$+\Omega$ and that well below has the same 
properties but is rotating with
rate $-\Omega$. Using the Shakura-Sunyaev 
alpha turbulence model, we find
self-similar solutions where a
thin (relative to the full disk
thickness) equatorial layer
accretes very rapidly, essentially at free-fall speed. 
As a result the accretion speed is much
larger than it would be for
an alpha disk rotating in one direction. 
Counter-rotating accretion disks
may be a transient stage in the formation of 
counter-rotating galaxies and
in the accretion of matter onto compact 
objects. 
\end{abstract}


\section{INTRODUCTION}
The widely considered models of accretion disks have gas rotating in one
direction with a turbulent viscosity acting to
transport angular momentum outward
(\cite{SS}). However, recent observations point to more complicated disk
structures in both active galactic nuclei and on a
galactic scale. Warped and tilted nuclear accretion disks have been detected for
example in NGC 4753 (\cite{TWIST}).
Recent high spectral resolution studies
of normal galaxies has revealed counter
rotating gas (\cite{CIRI}) and/or stars in many galaxies of all
morphological types -
ellipticals, spirals, and irregulars
(see reviews by Rubin 1994 and Galletta 1996). NGC 4826 (\cite{NGC4826})
and IC 1459 (\cite{IC1459}) are examples of galaxies with central
counter-rotating gas and stellar disks, respectively.
In elliptical galaxies, the
counter-rotating component is usually
in the nuclear core and may result from
merging of galaxies with opposite
angular momentum. Newly supplied gas
with misaligned angular momentum in
the nuclear region of a galaxy may
have important consequences for nuclear
activity if there is a rotating massive
black hole at the galaxy's center (\cite{SCH}). 
In contrast, in a number of spirals and
S0 galaxies, counter-rotating disks of
stars and/or gas have been found to
co-exist with the primary disk out
to large distances ($10 - 20$ kpc),
with the first example, NGC 4550, 
discovered by Rubin, Graham, and Kenney (1992).
It is not likely that the large scale
counter-rotating disks result from
mergers of flat galaxies with opposite
angular momenta because of the large
vertical thickening observed in
simulation studies of such mergers (\cite{BARNES}). 
Thakar and Ryden (1996) discuss
different possibilities, (a) that the
counter-rotating matter comes from the
merger of an oppositely rotating gas
rich dwarf galaxy with an existing
spiral, and (b) that the accretion
of gas occurs over the lifetime of
the galaxy with the more
recently accreted gas counter-rotating.
Subsequent star formation in the
counter-rotating gas may then
lead to counter-rotating stars. The two-stream instability between
counter-rotating gas and co-rotating stars may enhance the rate of gas
accretion (\cite{RVEL}).

\begin{figure}[htb]
\begin{center}
\leavevmode
\epsfysize=1.9in
\epsfbox{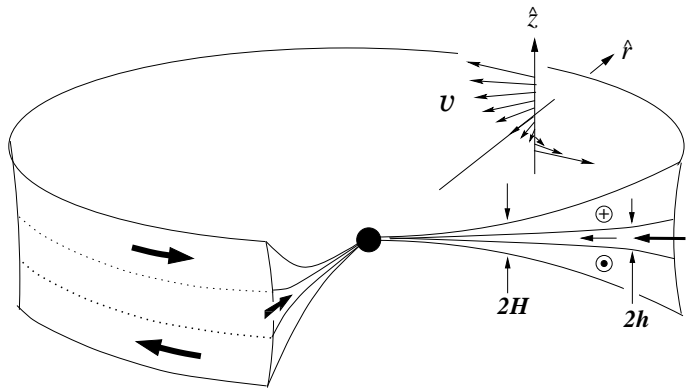}
\end{center}
\caption{Structure of two apposed, counter-rotating 
accretion disks and the midplane boundary layer. 
The inset shows the three dimensional view of the 
velocity field for the $n=1/2$ case shown in Figure 2. 
The velocity variation is analogous to that
in the Ekman layer of a rotating fluid (such as 
the ocean) where the Coriolis force balances the 
viscous force (\cite{GKB}).}
\label{FIG1}
\end{figure}

An important open problem is
how counter-rotating gas disks form and
what their structures are on galactic
scales and on the scale of disks in
active galactic nuclei. Here, we
investigate accretion disks consisting
of counter-rotating gaseous components
with gas at large $z$ rotating with
angular rate $+\Omega(r)$ and gas at
large negative $z$ rotating at rate
$-\Omega(r)$. The interface between the 
components at $z\!\sim\! 0$
constitutes a supersonic shear layer and 
is shown in Figure \ref{FIG1}.

\noindent Configurations of this type
may exist in astrophysical settings such as halo-disk interactions and the
accretion of newly supplied counter-rotating gas onto an existing
co-rotating disk. It might at first be supposed that powerful
Kelvin-Helmholtz instabilities heat
the gas to escape speed and rapidly destroy the assumed configuration. However,
supersonic shear layers exist and exhibit gross stability in stellar and
extra-galactic jets (\cite{HARDEE}).
In the counter-rotating disk, matter
approaching the equatorial plane from
above and below has angular momenta of
opposite signs with the result that
there is angular momentum annihilation
at $z=0$, the matter loses its centrifugal support and accretes at essentially
free-fall speed. On the other hand,
accretion disks rotating in one direction are modeled assuming a turbulent
viscosity
which is crucial for the outward
transport of angular momentum (\cite{SS}). The counter-rotating disks can
also be
expected to be turbulent owing in part
to the Kelvin-Helmholtz instability,
and turbulent viscosity can transport
angular momentum outward
in the large $|z|$ regions of the disk.

\section{THEORY AND SELF-SIMILAR SOLUTIONS}

For stationary, axi-symmetric disk-like
flows in cylindrical coordinates
$(r,\phi,z)$ with 

\begin{equation}
{\bf v} =
\left[v_r(r,z), v_\phi(r,z), v_z(r,z)\right],
\end{equation}

\noindent the momentum and continuity equations are

\begin{equation}
({\bf v}\cdot{\bf \nabla}) v_r
= {{v_\phi^2}\over r}
-{1\over \rho}{{\partial p}
\over{\partial r}}+{ g_r}+ \nu_{\!\perp} {\partial^{2} v_{z} \over \partial
z^2} +\nu_{\parallel}{\partial \over \partial r} \left[{\partial v_{r}
\over \partial r} +{v_{r} \over r}\right] 
\label{MOMR}
\end{equation}

\vspace{-2mm}

\begin{equation}
({\bf v}\cdot{\bf \nabla}) v_\phi =
{{-v_r v_\phi}\over r}
+\nu_{\!\perp}{\partial^{2}v_{\phi} \over \partial z^2}
+\nu_{\parallel}{\partial \over \partial r} \left[{\partial v_{\phi}\over
\partial r}+{v_{\phi}\over r}\right]
\label{MOMP}
\end{equation}

\vspace{-2mm}

\begin{equation}
({\bf v}\cdot{\bf \nabla}) v_z =
-{1\over\rho}{{\partial p}\over{\partial z}} +g_z
+\nu_{\parallel}\left[{\partial^{2}v_{z}\over \partial r^2} +{1 \over
r}{\partial v_{z} \over \partial r}\right] +\nu_{\!\perp}{\partial^{2}
v_{z} \over \partial z^2} \label{MOMZ}
\end{equation}

\vspace{-2mm}

\begin{equation}
{1\over r}{\partial\over {\partial r}}
\left( r \rho v_r \right) +{\partial\over \partial z} \left(\rho v_z\right) = 0.
\label{CONT}
\end{equation}

\noindent Here, $\rho(r,z)$ is the gas density, $p(r,z)$ is the pressure,
${\bf g} = -{\bf \nabla} \Phi $ is the gravitational acceleration with
$\Phi(r,z)$ the potential. For simplicity, terms
involving the gradient of the dynamic viscosity ($\rho \nu$) have been
neglected. Because the microscopic viscosity is negligible for the
conditions considered, we assume $\nu$ arises from small scale turbulence
and can be approximated by the ``alpha'' prescription of Shakura and
Sunyaev (1973) as $\nu = \alpha c_s H$, where $H$ is
the full half-thickness of the disk which is assumed thin ($H\ll r$), $c_s$
is the isothermal sound speed, and $\alpha$ is a dimensionless constant
much less than unity. Since the turbulent motions may not be isotropic,
we allow two different tensor components of $\nu$ ($\nu_{\parallel}$ for
shear in the $\hat{r}-\hat{\phi}$ direction and $\nu_{\perp}$ for shear in
the vertical direction) with possible different $r-$dependences. We also
assume the second viscosity
coefficient $\nu' = {2\over3} \nu$ so
that the stress tensor $T_{ij} = \rho v_i v_j + p\delta_{ij} + T^{\nu}_{ij}$
with the viscous contribution
$T_{ij}^{\nu} =-\nu (\partial v_i/\partial x_j + \partial v_j /\partial
x_i)$. Here, $\nu = \nu_{\!\perp}$ if $j = z$, and $\nu = \nu_{\parallel}$
otherwise, yielding equations (\ref{MOMR})-(\ref{MOMZ}).

We seek self-similar solutions to
equations (\ref{MOMR})-(\ref{CONT}) of the form

\begin{equation}
\begin{array}{c}
\displaystyle v_r(r,z)= -u_r(\zeta) V_c(r), \\[13pt]
\displaystyle v_\phi(r,z)= u_\phi(\zeta)V_c(r), \\[13pt]
\displaystyle v_z(r,z) = -\left({{h(r)}\over r}
\right)u_z(\zeta)V_c(r), \\[13pt] 
\displaystyle\rho(r,z)= \rho_0(r) Z
\left({z\over H(r)}\right), 
\label{SOLN}
\end{array}
\end{equation}

\noindent where $\zeta \equiv {z/ h(r)}$ is the dimensionless vertical
distance in the
disk with $h(r)$ a length scale identified subsequently, and $V_c(r) =
\left[ r({\partial \Phi}/ {\partial r})\vert_{z=0}\right]^{1/2}$
is the circular velocity of the gas.
The potential is due in general to disk
and halo matter and a central object.
Here and subsequently we neglect the pressure 
force in the radial equation of motion.
We consider cases where
$V_c(r) = (GM/r_{0})^{1/2}(r_{0}/r)^n $, 
with $r_0$ a constant length
scale. The value $n=1/2$ corresponds to a 
Keplerian disk around an object
of mass $M$, and $n=0$ to a flat rotation 
curve applicable to flat galaxies.

Substitution of equation (\ref{SOLN}) into equations (\ref{MOMR}) and
(\ref{MOMP}) gives

\begin{eqnarray}
(1+\epsilon^2\eta^2\zeta^2)u_r''
= u_\phi^2+nu_r^2-1+\eta\zeta u_r u_r'\,\,\,\,
\hspace{4mm}\, \nonumber \\
-u_zu_r'-\epsilon^2\zeta
\eta(\eta+2n)u_r' +\epsilon^2(1-n^2)u_r, 
\label{DIFFR}
\end{eqnarray}

\vspace{-3mm}

\begin{eqnarray}
(1+\epsilon^2\eta^2\zeta^2)u_\phi'' =
(n-1)u_r u_\phi-u_z u_\phi' \,\,\,\,\,\,\,
\hspace{7mm}\,\,
\nonumber \\
+\eta \zeta u_r u_\phi'-
\epsilon^2 \zeta\eta(\eta+2n)u_\phi'+
\epsilon^2(1-n^2)u_\phi, 
\label{DIFFP}
\end{eqnarray}

\noindent where a prime denotes a
derivative with respect to $\zeta$.
We have chosen the scaling such
that $\nu_{\perp} r/ (V_{c}h^{2}) \equiv 1$ and $\epsilon^2 \equiv
\nu_{\parallel} /r V_{c}$, and defined $\eta \equiv (r/h)(dh/dr)$.
We assume that the disk is not strongly
self-gravitating and that it is vertically isothermal so that its overall
half-thickness is $H \sim (c_s/V_c)r$
which is the same as for a disk rotating in one direction. From the
assumed scaling, $h^2 = \nu_{\perp} r/V_{c}$, and an assumed alpha viscosity
prescription $\nu_{\parallel} = \alpha c_{s}H$ and $\nu_{\!\perp} = \alpha
c_{s}h$ (\cite{REGEV}),
$\delta \equiv h/H \sim \alpha \ll 1$.
As a result, (\ref{MOMZ}) simplifies to
$(\partial_{z} p)= -\rho\partial_{z} \Phi$. which determines the isothermal
density profile $Z(\zeta) = \exp(-\delta^{2} \zeta^{2})$. From the assumed
scaling
$\rho_{0}(r) \propto r^{-\beta -1/2}$,
equation (\ref{CONT}),

\begin{eqnarray}
u_z'=\left[\beta+n-{1\over2}-2\delta^2\zeta^2 
\left({r \over H}{d H \over d r}\right)\right] 
u_r \nonumber \\
+\eta\zeta u_r' + 2\delta^{2}\zeta u_{z},
\label{DIFF4}
\end{eqnarray}

\noindent determines $v_{z}$. For consistency of the self-similar
solutions, $\epsilon$ must be independent of $r$, implying $\eta = 1$ and
$(r/H)(dH/dr) = 1$ if the viscosities scale similarly, $\nu_{\perp} \sim
\nu_{\parallel}$.

Equations (\ref{DIFFR})-(\ref{DIFF4})
constitute a closed system for the
dimensionless functions $u_r(\zeta), u_\phi(\zeta),$ 
and $u_z(\zeta)$. Applying equations
(\ref{DIFFR})-(\ref{DIFF4}) to the flow
suggested in Figure 1 \ref{FIG1},
we infer that $u_r(\zeta)$ is an even function 
of $\zeta$, while $u_\phi(\zeta)$ and $u_z(\zeta)$ 
are odd functions. For $\vert \zeta \vert \gg 1$, we
impose $u_r \simeq (1+n)\epsilon^2$ and $u_\phi 
\simeq \pm \sqrt{1-(n+1)^2\epsilon^4}$ so as to 
have $u_r'' \rightarrow 0$ and $u_\phi'' 
\rightarrow 0$. These limits correspond to an 
$\alpha$ disk rotating in one direction far away 
from the $\zeta = 0$ midplane. For
$n=1/2$, $u_r(\zeta \gg 1) \simeq 3\epsilon/2$ 
and $v_{r}(\zeta \gg 1) = 3\nu_{\parallel}/2r$, 
agreeing with the radial velocity of a single thin
Keplerian disk with viscosity $\nu_{\parallel}=
\epsilon^2 r V_{c}$.

A steady counter-rotating disk
may result from gas continuously
supplied at large $r$. Similarly, steady state 
solutions where gas is entrained
and delivered vertically onto the faces
of the disk may exist. We assume that
$v_{z}\rho(r,z) \rightarrow 0$ as
$\vert z \vert \rightarrow \infty$.
Considering only this subset of solutions, 
we demand the accretion rate

\begin{eqnarray}
\dot M = 2\pi\int_{-\infty}^{\infty} dz r 
\rho(r,z) \vert v_r(r,z)\vert\simeq 2\pi r
\rho_0(r)\nonumber \\
\times V_c(r)h(r) \int_{-H/h}^{H/h} d\zeta u_r(\zeta)
e^{-\delta^{2}\zeta^{2}},
\label{MDOT}
\end{eqnarray}

\noindent be a constant. A sufficient condition for this is
$\beta=\eta-n+1/2$ and $\delta = h/H$ is $r-$independent. For an alpha disk
rotating in one direction, $\dot M_{SS} = 2\pi r \Sigma |v_r|_{SS},$ with
the accretion speed $|v_r|_{SS} \sim \alpha c_s H/r \ll c_s$, where
$\Sigma=\int dz \rho$ is the disk's surface mass density (\cite{SS}).
In contrast, for a counter-rotating disk equation (\ref{MDOT}) gives $\dot
M_{CR} = 2\pi r\Sigma |v_r|_{cr}$, with accretion speed (averaged over $z$)
$|v_r|_{CR} \sim (h/H)V_c$. For the same $\alpha$, we have $|v_r|_{CR} \gg
|v_r|_{SS}$: the accretion speed of the counter-rotating disk is much
larger than that of the disk rotating in one direction.

A further constraint on solutions of
equations (\ref{MOMR})-(\ref{CONT}) is obtained by considering the energy
dissipation and radiation of the disk.
The viscous dissipation per unit
area of the disk is

\begin{eqnarray}
D(r) =\int_{-H}^H dz~\rho~\nu
\bigg[2\bigg({{\partial v_r}\over {\partial
r}}\bigg)^2+\bigg(r{{\partial \Omega}
\over{\partial r}}\bigg)^2 \nonumber \\
+2\bigg({{\partial v_z}\over {\partial z}}\bigg)^2 
+\bigg({{\partial v_z}\over {\partial
r}}+{{\partial v_r}\over {\partial z}}\bigg)^2+
\bigg({{\partial v_\phi}\over {\partial
z}}\bigg)^2\bigg],
\label{D}
\end{eqnarray}

\noindent where $\Omega = v_\phi/r$.
For a disk rotating in one direction, the dominant 
contribution to $D(r)$ is from the $\partial 
\Omega/\partial r$ term and this
gives $D_{SS}(r)\approx \Sigma \nu_{\parallel} 
(n+1)^2 (V_c/r)^2 = \dot M_{SS}(V_c/r)^2
(n+1)^2/(3\pi)$. In contrast, for a counter-rotating
disk, the dominant contributions are from the 
terms $\partial v_r/\partial
z$ and $\partial v_\phi/\partial z$ with the 
result 

\vspace{-3mm}
\begin{eqnarray}
D_{CR}(r) \simeq \Sigma \nu_{\!\perp} 
V_c^2/(hH)\,\,\,  \hspace{2.5cm} \,\,\,\nonumber \\
\sim  \dot M_{CR}
(V_c/r)^2/(3\pi) \sim (r/H)^2D_{SS}(r).
\end{eqnarray}
\vspace{-3mm}

\begin{figure}[htb]
\begin{center}
\leavevmode
\epsfxsize=2.8in
\epsfbox{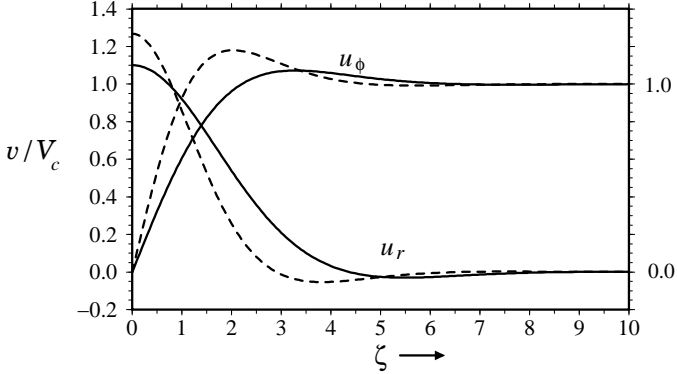}
\end{center}
\caption{Solutions to Equations
(\ref{MOMR}), (\ref{MOMP}), and (\ref{CONT}) 
for $\epsilon = 0.01$ and $\delta = 0.1$ as a 
function of $\zeta = z/h(r)$. Solid lines correspond to
a Keplerian disk ($n=1/2\,, \beta = 1$) with $u_{r}(0) = 1.10007597\ldots$
and $u_{\phi}^{\prime}(0) = 0.66260669\ldots$. The dashed curves correspond
to $n=0$ (flat rotation curve potential) and $\beta = 3/2$ and have initial
values $u_{r}(0) = 1.26728681248\ldots$ and $u_{\phi}^{\prime}(0) =
1.1256582571\ldots$. These values are close to those of the $\delta =
\epsilon = 0$ ``zero-order'' solutions: $u_{r}(0) =
[u_{\phi}^{\prime}(0)]^{2} = 1.100112673\ldots\,(n=1/2)$ and
$u_{r}(0) = [u_{\phi}^{\prime}(0)]^{2}
= 1.2672806222\ldots\, (n=0)$. All solutions reach their asymptotic values
for $\zeta \rightarrow \infty$ and the vertical velocity $v_{z}/V_{c}
\approx 0$ on this scale. Note that the midplane infall velocity for the
Keplerian case is
shear-reduced by a
factor $\sim 0.77$ from the free-fall speed $(2GM/r)^{1/2}$. }
\label{PLOT}
\end{figure}

\noindent The dissipated energy is radiated from the 
faces of the disk if it is
optically thick in the $z$-direction and 
thus $D=4acT^4/(3\kappa\Sigma)$,
where $a$ is the radiation constant, $c$ 
the speed of light, $\kappa$ the
opacity, and $T$ is the internal disk 
temperature. For the general scaling
$\kappa \propto \rho^a T^b$, the counter-rotating 
disk thus has $T \propto r^{-\xi}$ with $\xi = 
(3+2a+n)/(4-b+a/2)$ and $h \propto T^{1/2} r^{1+n} 
\propto r^\eta$. For example, for Kramer's
opacity $a=1,~b=-3.5$, we have $\xi = (5+n)/8$ 
and $\eta = (11+15n)/16$,
whereas for electrons scattering
opacity $a=0,~b=0$, and $\xi =(3+n)/4$ and 
$\eta=(11+9n)/8$. In general,
the strict requirement $\eta=1$ for our 
self-similar solutions will not be satisfied, but
this is not expected to be important if 
$\epsilon \ll 1$ (see Figure
\ref{PLOT})\footnote{ Because the viscosity 
scaling $\nu_{\perp} \sim \nu_{\parallel}$ 
implies $\eta = 1$, values of $\eta \neq 1$ 
are strictly consistent with our scalings 
and constant $\dot M$ only if
the turbulent viscosity is anisotropic such 
that shear perpendicular to the
plane is governed by $\nu_{\parallel} = 
\epsilon^2 rV_{c} \sim r^{1-n}$,
but shear in the plane is determined by
$\nu_{\perp} = h^{2}V_{c}/r \sim r^{2\eta-n-1}$.}. 
The apparent surface temperature of the disk 
$T_{{\em eff}}(r)$ is given by $D(r)=2\sigma
T_{{\em eff}}^4$, where $\sigma =ac/4$ is the 
Stefan-Boltzmann constant. The dependence for 
the counter-rotating disk, $T_{{\em eff}} \propto
r^{-(n+1)/2}$, is the same as for an alpha disk 
rotating in one direction. Similarly, the 
spectrum, $F_\omega$, obtained by integrating 
the Planck function $B_\omega(T_{{\em eff}}(r))$ 
over the surface area of the disk, $F_\omega \propto 
\omega^{(3n-1)/(n+1)}$, is the same as for a disk
rotating in one direction. For an optically thin 
disk, $D=2H\Lambda$, where $\Lambda(\rho, T)$ 
is the emissivity of the disk matter.

In the limit $\epsilon = \delta = 0$
and $\beta = \eta -n + 1/2$, equation (\ref{DIFF4}) gives $u_z = \eta \zeta
u_r$, which reduces equations (\ref{DIFFR}) and (\ref{DIFFP}) to
$u_r^{\prime\prime} = u_\phi^2 +nu_r^2-1$ and $u_\phi'' = (n-1)u_ru_\phi$,
respectively. These have the integral

\begin{eqnarray}
(u_\phi')^2-{1-n\over2}(u_r')^2+(1-n)u_r u_\phi^2 
\,\,\hspace{12mm}\,\, \nonumber \\
+{n(1-n)\over3}u_r^3
-(1-n)u_r = \mbox{const.}
\label{CONST}
\end{eqnarray}

\noindent Evaluating (\ref{CONST})
at $\zeta=0$ and $\zeta = \infty$ gives $[u_\phi'(0)]^2 = (1-n)u_r(0)
-n(1-n)u_r^{3}(0)/3$. We utilize this result by first finding the single
independent initial condition $u_{r}(0)$ for $\delta = \epsilon = 0$, then
allow $\delta\,,\epsilon \neq 0$, and perturbatively find the proper values
of $u_{r}(0)$ and $u_{\phi}^{\prime}(0)$. Three of five initial conditions
are fixed by symmetry: $u_{r}^{\prime}(0) = u_{\phi}(0) = u_{z}(0) = 0$. We
search for solutions by tuning $u_{r}(0)$ and $u_{\phi}^{\prime}(0)$,
regarding $\epsilon$ and $\delta$ as fixed. In principle, the required values
$u_{r}(\infty)$ and $u_{\phi}(\infty)$
are satisfied for a range of $\delta$.
The additional parameter $\delta \gg \epsilon$ can be chosen according to
the behavior of $u_{z}(\infty)$ desired. Thus, a class of solutions exist
located on curved segments in
$[u_{r}(0), u_{\phi}^{\prime}(0), \delta]$ space.

Figure \ref{PLOT} shows solutions to equations (\ref{DIFFR})-(\ref{DIFF4})
for $\epsilon = 0.01$ and $\delta = 0.1$, corresponding to $h/H \simeq
1/10$ and $h/r \simeq 1/300$. Both Keplerian ($n=1/2\,, \beta = 1$) and
galactic ($n=0\,, \beta = 3/2$) solutions are shown. The solutions are
sensitive to the precise values of $u_{r}(0)\,,
u_{\phi}^{\prime}(0)\,,\delta$, and $\epsilon$. Note that for $3 \lesssim
|\zeta| \lesssim 7$, the solutions have a region with $u_{r} < 0$
indicating spiraling outflows on both sides of the midplane. The vertical
velocities $v_{z}/V_{c} \ll u_{r}$ or $u_\phi$ are indistinguishable from
the $x$-axis on the scale of Fig. \ref{PLOT}. For large $\zeta$, the
density fall off
gives $\rho v_z \rightarrow 0$.

\section{DISCUSSION}

Counter-rotating gas supplied to the outer part of an existing
co-rotating gas disk will increase the mass accretion rate.
Comparing accretion rates of a standard $\alpha-$disk ($\dot{M}_{SS}$) with
that of a counter-rotating Keplerian disk ($\dot{M}_{CR}$) of the same
$\Sigma$ and $M$, we find from equation (\ref{MDOT}) $\dot{M}_{CR}/
\dot{M}_{SS}
\simeq 1 + 0.58(\delta /\epsilon^{2})$. For the values of Figure 2 this
ratio is $ \simeq 580$.

Thermal and/or dynamical instabilities may destroy
the counter-rotating accretion flows described above.
The relative importance of thermal and dynamical instabilities can be
estimated by comparing the thermal
dissipative time scale, $\tau_{Q} \approx \Sigma c_{s}^2/D_{CR}(r) \simeq
(r/h)(H/V_{c})(c_{s}/V_{c})^2$, with dynamical and viscous time scales,
$\tau_{z}\equiv h/c_{s},\,\tau_{\perp}\equiv r/V_{c}$, and
$\tau_{\nu}\equiv r^2/\nu_{\!\perp}$. We find $\tau_{Q}\sim
\epsilon^2\delta^{-3}\tau_{z} \sim \epsilon^2
\delta^{-5/2}\tau_{\!\perp}\sim \epsilon^{4} \delta \tau_{\nu}$. For
$\epsilon = 0.01$ and $\delta = 0.1$, the thermal dissipation time
$\tau_{Q} \ll \tau_{z} <\tau_{\!\perp}
\ll \tau_{\nu}$. In contrast
to the standard $\alpha-$disk, thermal
equilibrium is maintained on time scales of possible flow instabilities of the
inner shear layer.  The Kelvin-Helmholtz 
instabilities are likely to be the most important
(\cite{RAY} and \cite{CL}).  If the
counter-rotating disk is treated as a vortex sheet, 
then a local stability analysis indicates unstable 
warping for wave numbers $|k_\phi/k_r|<
\sqrt2(c_s/\Omega r) \ll 1$.

Accretion of counter rotating gas
by an existing co-rotating gas disk may be a transient stage in the
formation of counter-rotating galaxies and in the accretion of matter onto
rotating black holes in active galactic nuclei. We
find that newly supplied counter-rotating gas drags inward the old
co-rotating gas with an equal mass of old and new gas accreting rapidly.
Thus the old co-rotating gas may be entirely ``used up'' (dragged to the
center of a galaxy or into a black hole) if the mass of newly supplied gas
exceeds that of the old gas disk.
Accretion onto the faces of an existing
thin disk may not have the symmetry shown in Figures 
\ref{FIG1} and \ref{PLOT}.

\begin{figure}[htb]
\begin{center}
\leavevmode
\epsfysize=.82in
\epsfbox{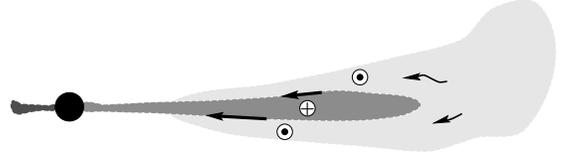}
\end{center}
\caption{Schematic drawing of accretion of newly supplied
counter-rotating gas ($\odot$) induced by viscous interaction with an
existing disk of co-rotating gas ($\oplus$). } 
\label{MERGER}
\end{figure}

\noindent There may instead be
two layers of rapid radial inflow bounding the existing gas disk near the
midplane as sketched in Figure \ref{MERGER}. Solutions
for this configuration can be composed from those of 
Figure \ref{PLOT} if $h \ll H$.

\vspace{5mm}
\acknowledgments

We thank M. Haynes and K. Jore for valuable discussions. T. C. acknowledges
support from NSF grant DMR-9300711. R.L. was supported in part by NASA
grant NAGW 2293 and NSF grant AST-9320068.

{}

\end{document}